\documentclass [a4paper,fleqn, 12pt]{article}
\usepackage{graphicx}
\usepackage[small]{subfigure,epsfig}

\usepackage {amsmath} \usepackage{amssymb} \usepackage{cite}

\newcommand{\cn}

\begin{document}


\title
{Vortices ans Polynomials:  Nonuniqueness of the Adler -- Moser polynomials for the Tkachenko equation}

\author
{Maria V Demina,  \and   Nikolay A. Kudryashov}

\date{Department of Applied Mathematics, National Research Nuclear University
MEPHI, 31 Kashirskoe Shosse,
115409 Moscow, Russian Federation}




\maketitle

\begin{abstract}

Stationary and translating relative equilibria of point vortices in the plane are studied. It is shown that stationary equilibria of a system containing point vortices with arbitrary choice of circulations can be described with the help of  the Tkachenko equation. It is obtained that the Adler -- Moser polynomial are not unique polynomial solutions of the Tkachenko equation. A generalization of the Tkachenko equation to the case of translating relative equilibria is derived. It is shown that the generalization of the Tkachenko equation possesses polynomial solutions with degrees that are not triangular numbers.

\end{abstract}






\section{Introduction}

The problem of finding stationary and relative equilibrium solutions to Helmholtz's equations describing motion of point vortices in the plane has been attracting much attention during recent years \cite{Aref01, Aref02, Aref03, Aref04, Oneil01, Oneil02, Oneil03, Clarkson01, Borisov01}.  The so--called "polynomial method" is often used while studying this problem \cite{Aref01}. According to this method one introduces polynomials with roots at vortex positions. For example, it was found by Stieltjes that the generating polynomial of $N$ identical point vortices on a line is essentially the $N$th Hermite polynomial.  Considering equilibrium of point vortices with equal in absolute value circulations leads to an ordinary differential equation first discovered by Tkachenko \cite{Aref01}. It is well known that the  Adler -- Moser polynomials  give polynomial solutions to the Tkachenko equation. Originally these polynomials arose in the theory of one of the most famous soliton  equation, the Korteweg -- de Vries equation \cite{Adler01}. This fact provides a remarkable and rather unexpected connection between the dynamics of point vortices and the theory of integrable partial differential equations \cite{Bartman01}. Not long ago analogous connections between equilibria of point vortices with circulations $\Gamma$,~$-2 \Gamma$ and rational solutions of the Sawada~-- Kotera and the  the Kaup -- Kupershmidt equations was established \cite{Demina25}.

In this article we study stationary and translating relative equilibria of multivortex systems with circulations $\Gamma_1$, $\ldots$, $\Gamma_N$. Our aim is to show that the stationary case can be described with the help of the Tkachenko equation and the translating case is reducible to a generalization of the Tkachenko equation. As a consequence of our results we obtain that the  Adler -- Moser polynomials are not unique polynomial solutions of the Tkachenko equation.

This article is organized as follows. In section \ref{Stationary_case} we consider stationary equilibria of point vortices with arbitrary choice of circulations. We derive an ordinary differential equation satisfied by generating polynomials of arrangements and transform this equation to the Tkachenko equation.   Section \ref{Translating_case} is devoted to the translating case. We find an ordinary differential equation satisfied by generating polynomials of the vortices and reduce the resulting equation to the generalization of the Tkachenko equation. In section \ref{Properties_Tkachenko} we study properties of the Tkachenko equation and the generalized Tkachenko equation. Quite unexpectedly it is possible to construct polynomial solutions of the Tkachenko equation not included into the sequence of the Adler -- Moser polynomials. We present these solutions in section \ref{Properties_Tkachenko}. Usually it is supposed that the generalized Tkachenko equation possesses polynomial solutions only with  degrees being triangular numbers. We give an example that contradicts this assumption.

\section{Stationary equilibria of point vortices} \label{Stationary_case}

The motion of $M$ point vortices with circulations (or strengths)
$\Gamma_k$ at positions $z_k$, $k=1$,~$\ldots$~,~$M$ is described by the following system of differential equations
\begin{equation}
\label{Motion_of_Vortices}\frac{d z_k^{*}}{d\,t}=\frac{1}{2\pi i}\sum_{j=1}^{M}{}^{'}\frac{\Gamma_j}{z_k-z_j},\quad k=1,\ldots, M.
\end{equation}
The prime in this expression  means that we exclude the case $j=k$ and the symbol~$\,^{*}$ stands for complex conjugation. First of all we shall study stationary vortex configurations. Thus we set $d z_k^{*}/ dt =0$. Let us consider a multivortex system with $M$  vortices of circulations $\Gamma_1$,~$\ldots$~,~$\Gamma_N$. Suppose vortices at positions $a_{1}^{(j)}$,~$\ldots$~,~$a_{l_j}^{(j)}$ have the circulation $\Gamma_j$, $j=1$,~$\ldots$~,~$N$. This yields $M=l_1$ $+$ $\ldots$ $+$ $l_N$.  A convenient tool for analyzing such a situation is to introduce polynomials with roots at the positions of vortices \cite{Aref01}. In other words we subdivide the vortices into groups according to the values of circulations and define the polynomials
\begin{equation}
\label{Polynoials_at_positions_of_vortices_multi}P_j(z)=\prod_{i=1}^{l_j}(z-a_{i}^{(j)}),\quad j=1, \ldots, N.
\end{equation}
Thus we see that roots of the polynomial $P_j(z)$ give positions of vortices with circulation $\Gamma_j$. Note that the polynomials $P_j(z)$, $j=1$,~$\ldots$~,~$N$ do not have multiple and common roots. From equations \eqref{Motion_of_Vortices} we find the system of algebraic relations
\begin{equation}
\begin{gathered}
\label{Relations_for_positions_of_vortices}\sum_{j=1}^{N}\sum_{i=1}^{l_j}{}^{'}\frac{\Gamma_j}{a_{i_0}^{(j_0)}-a_{i}^{(j)}}=0,\quad i_0=1,\ldots, l_{j_0}, \quad j_0=1,\ldots, N,
\end{gathered}
\end{equation}
where the case $(j_0,i_0)=(j,i)$ is excluded. Using properties of the logarithmic derivative, we obtain the following equalities
\begin{equation}
\label{Polynoials_at_positions_of_vortices_Derivatives}P_{j,z}=P_j\sum_{i=1}^{l_j}\frac{1}{z-a_{i}^{(j)}},\quad P_{j,zz}=2P_j\sum_{i=1}^{l_j}\sum_{k=1}^{l_j}{}^{'}\frac{1}{(z-a_{i}^{(j)})(a_{i}^{(j)}-a_{k}^{(j)})}.
\end{equation}
Now let $z$ tend to one of the roots of the polynomial $P_{j_0}(z)$. Calculating the limit $z \rightarrow a_{i_0}^{(j_0)}$ in the expression for $P_{j_0,zz}$ yields
\begin{equation}
\label{Derivatives_a_j}P_{j_0,zz}(a_{i_0}^{(j_0)})=2P_{j_0,z}(a_{i_0}^{(j_0)})\sum_{i=1}^{l_{j_0}}{}^{'}\frac{1}{a_{i_0}^{(j_0)}-a_{i}^{(j_0)}}.
\end{equation}
Using equalities \eqref{Relations_for_positions_of_vortices}, \eqref{Polynoials_at_positions_of_vortices_Derivatives}, we get the conditions
\begin{equation}
\label{Conditions_at_roots}\Gamma_{j_0}\frac{P_{j_0,zz}(a_{i_0}^{(j_0)})}{P_{j_0,z}(a_{i_0}^{(j_0)})}=-2\sum_{j=1,\,j\neq j_0}^{N}\Gamma_j \frac{P_{j,z}(a_{i_0}^{(j_0)})}{P_{j}(a_{i_0}^{(j_0)})}, \quad i_0=1,\ldots, l_{j_0}, \quad j_0=1,\ldots, N,
\end{equation}
which are valid for any root $a_{i_0}^{(j_0)}$  of the polynomial $P_{j_0}(z)$. Further we see that the polynomial
\begin{equation}
\label{Polynomial_for_vort}\prod_{i=1}^{N}P_i(z)\left\{\sum_{j=1}^{N}\Gamma_j^2\frac{P_{j,zz}}{P_j}+2\sum_{k<j}\Gamma_k\Gamma_j\frac{P_{k,z}}{P_k}\frac{P_{j,z}}{P_j}\right\}
\end{equation}
 being of degree $M-2$ possesses $M$ roots $a_{i}^{(j)}$, $i=1$,~$\ldots$~,~$l_j$, $j=1$,~$\ldots$~,~$N$. Thus this polynomial identically equals zero. Consequently, the generating polynomials $P_j(z)$, $j=1$,~$\ldots$~,~$N$ of the arrangements described above satisfy the differential correlation
\begin{equation}
\label{Correlation_for_Polynomials}\prod_{i=1}^{N}P_i(z)\left\{\sum_{j=1}^{N}\Gamma_j^2\frac{P_{j,zz}}{P_j}+2\sum_{k<j}\Gamma_k\Gamma_j\frac{P_{k,z}}{P_k}\frac{P_{j,z}}{P_j}\right\}=0.
\end{equation}
Note that the reverse result is also valid. If $N$ polynomials $P_j(z)$, $j=1$,~$\ldots$~,~$N$ with no common and multiple roots satisfy correlation \eqref{Correlation_for_Polynomials}, then the roots of these polynomials give positions of vortices with circulations $\Gamma_j$, $j=1$,~$\ldots$~,~$N$ in stationary equilibrium.

Equation \eqref{Correlation_for_Polynomials} is invariant under the transformation $z\mapsto \alpha z+\beta$ with $\alpha$, $\beta$ being complex numbers and $\alpha\neq0$. This transformation converts any equilibrium arrangement into another equilibrium arrangement. We shall regard all such arrangements as equivalent.

Now we shall apply our results to the following multivortex situation. Suppose $l_n^{+}$ vortices with circulations $n\Gamma$ are situated at positions $z=a_{i}^{(n)}$, $i=1$, $\ldots$, $l_n^{+}$, $n=1$, $\ldots$, $N_1$ and $l_m^{-}$ vortices with circulations $-m\Gamma$ are situated at positions $z=b_{k}^{(m)}$, $k=1$, $\ldots$, $l_m^{-}$, $m=1$, $\ldots$, $N_2$. Further we consider the polynomials
\begin{equation}\begin{gathered}
\label{Polynoials_at_positions_of_vortices_PQ}P_n(z)=\prod_{i=1}^{l_n^{+}}(z-a_{i}^{(n)}),\quad n=1,\ldots , N_1 \hfill \\
 Q_m(z)=\prod_{k=1}^{l_m^{-}}(z-b_{k}^{(m)}),\quad m=1,\ldots , N_2
\end{gathered}
\end{equation}
with no common and multiple roots. The amount of vortices in such an arrangement is equal to $M=l_1^{+}+$ $\ldots$ $+$ $l_{N_1}^{+}$ $+$ $l_1^{-}$ $+$ $\ldots$ $+$ $l_{N_2}^{-}$. With the help of equation \eqref{Correlation_for_Polynomials}, we obtain
\begin{equation}
\begin{gathered}
\label{Correlation_for_Polynomials_Int}\prod_{n=1}^{N_1}P_n(z)\prod_{m=1}^{N_2}Q_m(z)\left\{\sum_{n=1}^{N_1}n^2\frac{P_{n,zz}}{P_n}+\sum_{m=1}^{N_2}m^2\frac{Q_{m,zz}}{Q_m} +2\sum_{n<j}nj\frac{P_{n,z}}{P_n}\frac{P_{j,z}}{P_j}\right.\\
\left.+2\sum_{m<k}mk\frac{Q_{m,z}}{Q_m}\frac{Q_{k,z}}{Q_k}-2\sum_{n,\, m}nm\frac{P_{n,z}}{P_n}\frac{Q_{m,z}}{Q_m}\right\}=0.
\end{gathered}
\end{equation}
Introducing new polynomials $\tilde{P}(z)$, $\tilde{Q}(z)$ according to the rules
\begin{equation}
\begin{gathered}
\label{New_Pols}\tilde{P}(z)=\prod_{n=1}^{N_1}P_n^{\frac{n(n+1)}{2}}(z)\prod_{m=1}^{N_2}Q_m^{\frac{m(m-1)}{2}}(z),\quad \tilde{Q}(z)=\prod_{n=1}^{N_1}P_n^{\frac{n(n-1)}{2}}(z)\prod_{m=1}^{N_2}Q_m^{\frac{m(m+1)}{2}}(z).
\end{gathered}
\end{equation}
and using the equalities
\begin{equation}
\begin{gathered}
\label{Relations_Tilde}\frac{d^2}{dz^2}\ln\left\{\tilde{P}(z)\tilde{Q}(z)\right\}=\sum_{n=1}^{N_1}n^2\left(\frac{P_{n,zz}}{P_n}-\frac{P_{n,z}^2}{P_n^2}\right)+ \sum_{m=1}^{N_2}m^2\left(\frac{Q_{m,zz}}{Q_m}-\frac{Q_{m,z}^2}{Q_m^2}\right), \hfill \\
\frac{d}{dz}\ln\left\{\frac{\tilde{P}(z)}{\tilde{Q}(z)}\right\}=\sum_{n=1}^{N_1}n\frac{P_{n,z}}{P_n}-\sum_{m=1}^{N_2}m\frac{Q_{m,z}}{Q_m}, \hfill
\end{gathered}
\end{equation}
we get the following differential equation for the polynomials $\tilde{P}(z)$ and $\tilde{Q}(z)$
\begin{equation}
\label{Equation_PQ_1}\tilde{P}_{zz}\tilde{Q}-2\tilde{P}_z\tilde{Q}_z+\tilde{P}\tilde{Q}_{zz}=0.
\end{equation}
This equation was in details studied by Burchnall and Chaundy \cite{Burchnall01}. In the framework of the vortex theory equation \eqref{Equation_PQ_1} is known as the Tkachenko equation \cite{Aref01}. Later this equation  appeared in the theory of the Korteweg -- de Vries equation \cite{Adler01}. Note that equation \eqref{Equation_PQ_1}  can de rewritten as
\begin{equation}
\label{Equation_PQ_1_Alt}\frac{d^2}{dz^2}\ln\left\{\tilde{P}(z)\tilde{Q}(z)\right\}+\left(\frac{d}{dz}\ln\left\{\frac{\tilde{P}(z)}{\tilde{Q}(z)}\right\}\right)^2=0.
\end{equation}
Along with this we see that substituting  equalities \eqref{Relations_Tilde} into this expression, we return back to  equation \eqref{Correlation_for_Polynomials_Int}.
Consequently, any polynomial solution of equation \eqref{Equation_PQ_1} of the form \eqref{New_Pols} describes stationary equilibrium of $M$ point vortices with circulations $\Gamma$, $2\Gamma$,~$\ldots$~,~$N_1\Gamma$ and $-\Gamma$, $-2\Gamma$,~$\ldots$~,~$-N_2\Gamma$. If the polynomial $\tilde{P}(z)$ has a root of multiplicity $n(n+1)/2$ at the point $z=z_0$ and the polynomial $\tilde{Q}(z)$ has a root of multiplicity $n(n-1)/2$ at the point $z=z_0$, then a vortex with circulation $n\Gamma$ is situated at the point $z=z_0$. Analogously, a vortex with circulation $-m\Gamma$ is situated at the point $z=z_0$ whenever  the polynomial $\tilde{P}(z)$ has a root of multiplicity $m(m-1)/2$ at the point $z=z_0$ and the polynomial $\tilde{Q}(z)$ has a root of multiplicity $m(m+1)/2$ at the point $z=z_0$.

Further we return to the multivortex situation we have studied in the beginning of this section. We would like to note that an equilibrium in the arrangements of vortices specified above necessarily exists if the following condition is valid
\begin{equation}
\begin{gathered}
\label{Condition_stationary}\sum_{j=1}^{N}l_j\Gamma_j^2-\left\{\sum_{j=1}^{N}l_j\Gamma_j\right\}^2=0
\end{gathered}
\end{equation}
This correlation can be obtained finding the Laurent series in a neighborhood of infinity for the expression in \eqref{Polynomial_for_vort} and setting the highest--order coefficient to zero. Let us show that we can reduce the equation \eqref{Correlation_for_Polynomials} satisfied by the generating polynomials $P_j(z)$, $j=1$,~$\ldots$~,~$N$ of the arrangements to equation  \eqref{Equation_PQ_1}.
Indeed, applying the transformation
\begin{equation}
\begin{gathered}
\label{New_Pols_Arbitrary_Circ}\tilde{P}(z)=\prod_{j=1}^{N}P_j^{\frac{\Gamma_j(\Gamma_j+1)}{2}}(z),\quad \tilde{Q}(z)=\prod_{j=1}^{N}P_j^{\frac{\Gamma_j(\Gamma_j-1)}{2}}(z).
\end{gathered}
\end{equation}
we obtain equation  \eqref{Equation_PQ_1}. We would like to note that the functions $\tilde{P}(z)$, $\tilde{Q}(z)$ may be algebraic now.

Thus we can describe stationary equilibria of $M$ point vortices with circulations $\Gamma_j$, $j=1$, $\ldots$, $N$ by means of equation \eqref{Equation_PQ_1}. Suppose we have found a solution of equation \eqref{Equation_PQ_1} in the form \eqref{New_Pols_Arbitrary_Circ}, then  a vortex with circulation $\Gamma_j$ is situated at the point $z=z_0$ whenever  the function $\tilde{P}(z)$ has a "root" of "multiplicity" $\Gamma_j(\Gamma_j+1)/2$ at the point $z=z_0$ and the function $\tilde{Q}(z)$ has a "root" of "multiplicity" $\Gamma_j(\Gamma_j-1)/2$ at the point $z=z_0$. The circulation is calculated as the difference of the corresponding "multiplicities". Here and in what follows we say that the function $f(z)$ has a "root" of "multiplicity" $k$ at the point $z=z_0$ if $f(z)=(z-z_0)^k \psi(z)$ with $\psi(z)$ being analytic in a neighborhood of $z_0$ and $\psi(z_0)\neq 0$. If $k \in \mathbb{N}$, then the point $z=z_0$ is a root of the function $f(z)$ in the usual cense.

Now let us consider the equation
\begin{equation}
\label{Equation_PQ_mu}S_{zz}T-2\mu S_zT_z+\mu^2ST_{zz}=0.
\end{equation}
The polynomials $S(z)$, $T(z)$ with no common and multiple roots that satisfy equation \eqref{Equation_PQ_mu} describe stationary equilibria of vortices with circulations $\Gamma$ and $-\mu \Gamma$, $\mu>0$. Roots of the polynomial $S(z)$ give positions of vortices with circulation $\Gamma$ and roots of the polynomial $T(z)$ give positions of vortices with circulation $-\mu \Gamma$. The case $\mu=2$ was studied in details in the articles \cite{Demina25, Loutsenko01}. Further we see that the transformations
\begin{equation}
\begin{gathered}
\label{New_Pols_mu}S(z)=\tilde{P}(z)\tilde{Q}(z)^{-\frac{\mu-1}{\mu+1}},\quad T(z)=\tilde{Q}(z)^{\frac{2}{\mu(\mu+1)}}\hfill \\
\tilde{P}(z)=S(z)T(z)^{\frac{\mu(\mu-1)}{2}},\quad \tilde{Q}(z)=T(z)^{\frac{\mu(\mu+1)}{2}}
\end{gathered}
\end{equation}
relate solutions of equations \eqref{Equation_PQ_1} and \eqref{Equation_PQ_mu}. Substituting expression \eqref{New_Pols_Arbitrary_Circ} into the first equality in \eqref{New_Pols_mu}, we obtain the relation
\begin{equation}
\begin{gathered}
\label{New_Pols_Arbitrary_Circ_mu}S(z)=\prod_{j=1}^{N}P_j^{\frac{\Gamma_j(\Gamma_j+\mu)}{\mu+1}}(z),\quad T(z)=\prod_{j=1}^{N}P_j^{\frac{\Gamma_j(\Gamma_j-1)}{\mu(\mu+1)}}(z).
\end{gathered}
\end{equation}
Thus we obtain that stationary equilibria of the multivortex system with circulations $\Gamma_j$, $j=1$, $\ldots$, $N$ can be described with the help of equation  \eqref{Equation_PQ_mu}. Suppose we have found a solution of equation \eqref{Equation_PQ_mu} in the form \eqref{New_Pols_Arbitrary_Circ_mu}, then  a vortex with circulation $\Gamma_j$ is situated at the point $z=z_0$ whenever  the point $z=z_0$ is a "root" of the function $S(z)$ with "multiplicity" $\Gamma_j(\Gamma_j+\mu)/(\mu+1)$ and the point $z=z_0$ is a "root" of the function $T(z)$ with "multiplicity" $\Gamma_j(\Gamma_j-1)/\{\mu(\mu+1)\}$. In order to find the circulation of the vortex at the point $z=z_0$ we take the "multiplicity" of the "root" $z=z_0$ of the function $S(z)$ and subtract $\mu$ multiplied by the "multiplicity"  of the "root" $z=z_0$ of the function $T(z)$.

Properties of equation \eqref{Equation_PQ_1} and some applications of our results we shall discuss in section \ref{Properties_Tkachenko}.

\section{Translating equilibria of point vortices} \label{Translating_case}

In this section we study the system of $M$ point vortices  moving uniformly with equal velocities. This motion is usually referred to as translating equilibria of point vortices.  Setting $d z_k/ dt =v^{*}$ with $v^{*}$ being a constant $v^{*}=\lambda/(2\pi i)$, $\lambda\neq 0$ in equations of motion \eqref{Motion_of_Vortices}, we obtain
\begin{equation}
\begin{gathered}
\label{Relations_for_positions_of_vortices_Translating}\sum_{j=1}^{M}{}^{'}\frac{\Gamma_j}{z_k-z_j}-\lambda=0,\quad k=1,\ldots, M.
\end{gathered}
\end{equation}
Again we subdivide the vortices into groups according to the values of circulations $\Gamma_j$, $j=1$,~$\ldots$~,~$N$. By $a_{1}^{(j)}$,~$\ldots$~,~$a_{l_j}^{(j)}$, $j=1$,~$\ldots$~,~$N$ we denote instantaneous positions of the vortices. We rewrite relations \eqref{Relations_for_positions_of_vortices_Translating} as follows
\begin{equation}
\begin{gathered}
\label{Relations_for_positions_of_vortices_Translating1}\sum_{j=1}^{N}\sum_{i=1}^{l_j}{}^{'}\frac{\Gamma_j}{a_{i_0}^{(j_0)}-a_{i}^{(j)}}-\lambda=0,\quad i_0=1,\ldots, l_{j_0}, \quad j_0=1,\ldots, N,
\end{gathered}
\end{equation}
Introducing the polynomials $P_j(z)$, $j=1$,~$\ldots$~,~$N$ with no multiple and common roots as given in formula \eqref{Polynoials_at_positions_of_vortices_multi}, we obtain the correlations
\begin{equation}
\label{Conditions_at_roots_Translating}\Gamma_{j_0}\frac{P_{j_0,zz}(a_{i_0}^{(j_0)})}{P_{j_0,z}(a_{i_0}^{(j_0)})}=-2\sum_{j=1,\,j\neq j_0}^{N}\Gamma_j \frac{P_{j,z}(a_{i_0}^{(j_0)})}{P_{j}(a_{i_0}^{(j_0)})}+2\lambda, \quad i_0=1,\ldots, l_{j_0}, \quad j_0=1,\ldots, N
\end{equation}
valid for any root $a_{i_0}^{(j_0)}$  of the polynomial $P_{j_0}(z)$.
Let us consider the polynomial
\begin{equation}
\label{Polynomial_for_vort_Translating}\prod_{i=1}^{N}P_i(z)\left\{\sum_{j=1}^{N}\Gamma_j^2\frac{P_{j,zz}}{P_j}+2\sum_{k<j}\Gamma_k\Gamma_j\frac{P_{k,z}}{P_k}\frac{P_{j,z}}{P_j}- 2\lambda\sum_{j=1}^{N}\Gamma_j\frac{P_{j,z}}{P_j} \right\}.
\end{equation}
This polynomial is of degree $M-1$ and possesses $M$ roots $a_{i}^{(j)}$, $i=1$,~$\ldots$~,~$l_j$, $j=1$,~$\ldots$~,~$N$. Thus this polynomial identically equals zero. Consequently the generating polynomials $P_j(z)$, $j=1$,~$\ldots$~,~$N$ of the arrangements described above satisfy the differential correlation
\begin{equation}
\label{Correlation_for_Polynomials_Translating}\prod_{i=1}^{N}P_i(z)\left\{\sum_{j=1}^{N}\Gamma_j^2\frac{P_{j,zz}}{P_j}+2\sum_{k<j}\Gamma_k\Gamma_j\frac{P_{k,z}}{P_k}\frac{P_{j,z}}{P_j}- 2\lambda\sum_{j=1}^{N}\Gamma_j\frac{P_{j,z}}{P_j}\right\}=0.
\end{equation}
The reverse result is also valid. If $N$ polynomials $P_j(z)$, $j=1$,~$\ldots$~,~$N$ with no common and multiple roots satisfy correlation \eqref{Correlation_for_Polynomials_Translating}, then the roots of these polynomials give instantaneous positions of vortices with circulations $\Gamma_j$, $j=1$,~$\ldots$~,~$N$ in translating equilibrium. Finding the Laurent series in a neighborhood of infinity for expression \eqref{Polynomial_for_vort_Translating} and setting the highest--order coefficient to zero, we obtain the following necessary condition
\begin{equation}
\label{Condition_Translating}\sum_{j=1}^{N}l_j\Gamma_j=0.
\end{equation}
for a translating equilibrium to exist. The transformation \eqref{New_Pols_Arbitrary_Circ} reduces equation \eqref{Correlation_for_Polynomials_Translating} to the equation
\begin{equation}
\label{Equation_PQ_1_Translating}\tilde{P}_{zz}\tilde{Q}-2\tilde{P}_z\tilde{Q}_z+\tilde{P}\tilde{Q}_{zz}-2\lambda\left(\tilde{P}_z\tilde{Q}-\tilde{P}\tilde{Q}_z\right)=0,
\end{equation}
which is a generalization of the Tkachenko equation \eqref{Equation_PQ_1}. This equation can be rewritten in the form
\begin{equation}
\label{Equation_PQ_1_Translating_Alt}\frac{d^2}{dz^2}\ln\left\{\tilde{P}(z)\tilde{Q}(z)\right\}+\left(\frac{d}{dz}\ln\left\{\frac{\tilde{P}(z)}{\tilde{Q}(z)}\right\}\right)^2 -2\lambda \frac{d}{dz}\ln\left\{\frac{\tilde{P}(z)}{\tilde{Q}(z)}\right\}=0.
\end{equation}
Consequently, we see that translating equilibria of $M$ point vortices with circulations $\Gamma_j$, $j=1$,~$\ldots$~,~$N$ can be described with the help of equation \eqref{Equation_PQ_1_Translating}.

\section{Properties of the Tkachenko equation and the generalized Tkachenko equation} \label{Properties_Tkachenko}

In this section we shall study properties of equations \eqref{Equation_PQ_1}, \eqref{Equation_PQ_1_Translating}. Suppose a pair $\tilde{P}(z)$, $\tilde{Q}(z)$ is a polynomial solution of equation \eqref{Equation_PQ_1}. Balancing the highest--order terms in equation \eqref{Equation_PQ_1}, we see that the degrees of the polynomials $\tilde{P}(z)$, $\tilde{Q}(z)$ are two successive triangular numbers. In other words $\deg \tilde{P}(z)=m(m+1)/2$, $\deg \tilde{Q}(z)=m(m-1)/2$, $m\in \mathbb{Z}$.

The Adler -- Moser polynomials provide a sequence of polynomial solutions to the Tkachenko equation. These polynomials can be constructed with the help of the following recurrence relation
\begin{equation}
\label{AM_Pols_construction}V_{k+1}(z)=V_{k-1}(z)\int\sigma_{k+1}\frac{V_k^{2}(z)}
{V_{k-1}^{2}(z)}dz,\quad k\in \mathbb{N}.
\end{equation}
The sequence begins with $V_0(z)=1$, $V_1(z)=z+c_1$. Usually the parameter $c_1$ is taken as zero. This choice results from the invariance of the Tkachenko equation under the transformation $z\mapsto \alpha z+\beta$. At each step an arbitrary constant $c_{k+1}$ is introduced in \eqref{AM_Pols_construction} and the parameter $\sigma_{k+1}$ is chosen in such a way that the corresponding polynomial is monic.  Another way to find the Adler -- Moser polynomials in explicit form is to use the Darboux transformations of the operator $d^2/dz^2$. As a result the Wronskian representation arises. Some other recurrence relations satisfied by these polynomials are given in \cite{Adler01, Kudr08a, Demina15, Kudr08, Kudr07}. Two successive Adler -- Moser polynomials solve the Tkachenko equation, i.~e. we may set $\tilde{P}(z)=V_{k\pm 1}(z)$, $\tilde{Q}(z)=V_{k}(z)$ in \eqref{Equation_PQ_1}. As a consequence of our results (see section \ref{Stationary_case}) we can obtain new solutions of equation  \eqref{Equation_PQ_1}. Suppose a pair $P(z)$, $Q(z)$  provides a solution of the Tkachenko equation, then the functions
\begin{equation}
\begin{gathered}
\label{New_Pols_Sol_to_AM}\tilde{P}(z)=P^{\frac{\Gamma(\Gamma+1)}{2}}(z)Q^{\frac{\Gamma(\Gamma-1)}{2}}(z),\quad \tilde{Q}(z)=P^{\frac{\Gamma(\Gamma-1)}{2}}(z)
Q^{\frac{\Gamma(\Gamma+1)}{2}}(z).
\end{gathered}
\end{equation}
also satisfy equation \eqref{Equation_PQ_1}. In particular we obtain that the functions
\begin{equation}
\begin{gathered}
\label{New_Pols_Sol_to_AM_Ex}\tilde{P}_k^{(1)}(z)=V_k^{\frac{\Gamma(\Gamma+1)}{2}}(z)V_{k+1}^{\frac{\Gamma(\Gamma-1)}{2}}(z),\quad \tilde{Q}_k^{(1)}(z)=V_k^{\frac{\Gamma(\Gamma-1)}{2}}(z)
V_{k+1}^{\frac{\Gamma(\Gamma+1)}{2}}(z), \quad k\in \mathbb{N}\cup \{0\}
\end{gathered}
\end{equation}
solve equation \eqref{Equation_PQ_1}. If $\Gamma \in \mathbb{Z}$, then $\tilde{P}_k^{(1)}(z)$, $\tilde{Q}_k^{(1)}(z)$ in \eqref{New_Pols_Sol_to_AM_Ex} are polynomials.

Further let us consider an example. Suppose we study stationary equilibria of $l_1^{+}$ point vortices with circulation $\Gamma$ and $l_2^{-}$ point vortices with circulation $-2\Gamma$. In our situation we have only two groups of vortices, thus we shall omit corresponding indices.  In other words we set $S(z)\equiv P_1(z)$, $Q_1(z)=1$, $T(z)\equiv Q_2(z)$ (see \eqref{Polynoials_at_positions_of_vortices_PQ}). According to the formula \eqref{Correlation_for_Polynomials_Int} the polynomials
\begin{equation}\begin{gathered}
\label{Polynoials_at_positions_of_vortices_PQ_mu_2}S(z)=\prod_{i=1}^{l_n^{+}}(z-a_{i}),\quad
 T(z)=\prod_{k=1}^{l_2^{-}}(z-b_{k})
\end{gathered}
\end{equation}
describing these configurations satisfy the equation
\begin{equation}
\label{Correlation_for_Polynomials_mu_2}S_{zz}T-4 S_zT_z+4ST_{zz}=0.
\end{equation}
Note that the vortices with circulation $\Gamma$ are situated at the points $a_i$, $i=1$,~$\ldots$~,~$l_1^{+}$ and  the vortices with circulation $-2\Gamma$ are situated at the points $b_k$, $k=1$,~$\ldots$~,~$l_2^{-}$. In new variables
\begin{equation}
\begin{gathered}
\label{New_Pols_mu_2}\tilde{P}(z)=S(z)T(z),\quad \tilde{Q}(z)=T^3(z).
\end{gathered}
\end{equation}
we rewrite equation \eqref{Correlation_for_Polynomials_mu_2} in the form \eqref{Equation_PQ_1}. In order to obtain relation \eqref{New_Pols_mu_2} we used expression \eqref{New_Pols}. Polynomial solutions of  equation \eqref{Correlation_for_Polynomials_mu_2} were studied in  \cite{Demina25, Loutsenko01}. Two neighbor polynomials from the sequences
\begin{equation}
\begin{gathered}
\label{Vortices_Polynomials_Solution_mu}S_{k+1}(z)=S_{k-1}(z)\int\gamma_{k+1}\frac{T_k^{4}(z)}{S_{k-1}^{2}(z)}dz,\quad k\in \mathbb{N} \hfill \\
T_{k+1}(z)=T_{k-1}(z)\int\delta_{k+1}\frac{S_k^{}(z)}{T_{k-1}^{2}(z)}dz,\quad k\in \mathbb{N}. \hfill
\end{gathered}
\end{equation}
i.~e. $S_k$, $T_{k+1}$ and  $S_k$, $T_{k-1}$, provide polynomial solutions of equation \eqref{Correlation_for_Polynomials_mu_2}. The "initial conditions" for these sequences are $S_{0}(z)=1$, $S_{1}(z)=z+s_1$, $T_{0}(z)=1$, $T_{1}(z)=z+t_1$. Again we can set $s_1=0$, $t_1=0$. Calculating the indefinite integrals in expressions \eqref{Vortices_Polynomials_Solution_mu}, we introduce the integration constant $s_{k+1}$ in the first one and $t_{k+1}$ in the second. For convenience the parameters $\gamma_{k+1}$, $\delta_{k+1}$ can be chosen in such a way that all the polynomials are monic. Consequently, we obtain that the Tkachenko equation admits polynomial solutions of the form
\begin{equation}
\begin{gathered}
\label{New_Pols_mu_2_EX}\tilde{P}_k^{(2)}(z)=S_{k}(z)T_{k+1}(z),\quad \tilde{Q}_k^{(2)}(z)=T_{k+1}^3(z),\quad k\in \mathbb{N} \cup \{0\},\hfill \\
\tilde{P}_k^{(3)}(z)=S_{k}(z)T_{k-1}(z),\quad \tilde{Q}_k^{(3)}(z)=T_{k-1}^3(z),\quad k\in \mathbb{N} \hfill
\end{gathered}
\end{equation}
with $S_{k}(z)$, $T_{k\pm 1}(z)$ given by \eqref{Vortices_Polynomials_Solution_mu}, see also table \ref{T:PQ_Tild}.

\begin{table}[b]
    \caption{Polynomials from the sequences \eqref{New_Pols_mu_2_EX}.} \label{T:PQ_Tild}
       \begin{tabular}[pos]{|l|l|}
                \hline
                $\tilde{P}_0^{(2)}(z) = z$ & $\tilde{Q}_0^{(2)}(z) = z^3$\\
$\tilde{P}_1^{(2)}(z) = z^3+t_2z$ & $ \tilde{Q}_1^{(2)}(z)={z}^{6}+3\,t_{{2}}{z}^{4}+3\,{t_{{2}}}^{2}{z}^{2}+{t_{{2}}}^{3}$\\
$\tilde{P}_2^{(2)}(z) = {z}^{10}+t_{{3}}{z}^{6}-3s_{{2}}{z}^{5}+s_{{2}}t_{{3}}z-4s_2^2$ & $ \tilde{Q}_2^{(2)}(z)={z}^{15}+3t_{{3}}{z}^{11}-12\,s_{{2}}{z}^{10}+$\\
$ $ & $\qquad \quad+3t_3^2{z}^{7
}-24s_{{2}}t_{{3}}{z}^{6}+48s_2^2{z}^{5}+t_3^3{z}^
{3}-$\\
$ $ & $\qquad \quad-12t_3^2s_{{2}}{z}^{2}+48s_2^2t_{{3}}z-64s_2^3$\\
\hline
 $\tilde{P}_1^{(3)}(z) = z$ & $\tilde{Q}_1^{(3)}(z) = 1$\\
$\tilde{P}_2^{(3)}(z) = z^6+s_2z$ & $ \tilde{Q}_2^{(3)}(z)=z^3$\\
$\tilde{P}_3^{(3)}(z) = {z}^{10}+{\frac {33}{5}}\,t_{{2}}{z}^{8}+{\frac {98}{5}}\,{t_{{2}}}^{2
}{z}^{6}+42\,{t_{{2}}}^{3}{z}^{4}$ & $ \tilde{Q}_3^{(3)}(z)={z}^{6}+3t_{{2}}{z}^{4}+3t_2^2{z}^{2}+t_2^3$\\
$\qquad  \qquad +s_{{3}}{z}^{3}+21t_2^4{z}^
{2}+s_{{3}}t_{{2}}z-7t_2^5$ & $ $\\
\hline
        \end{tabular}
\end{table}

\begin{figure}[t]
 \centerline{
 \subfigure[]{\epsfig{file=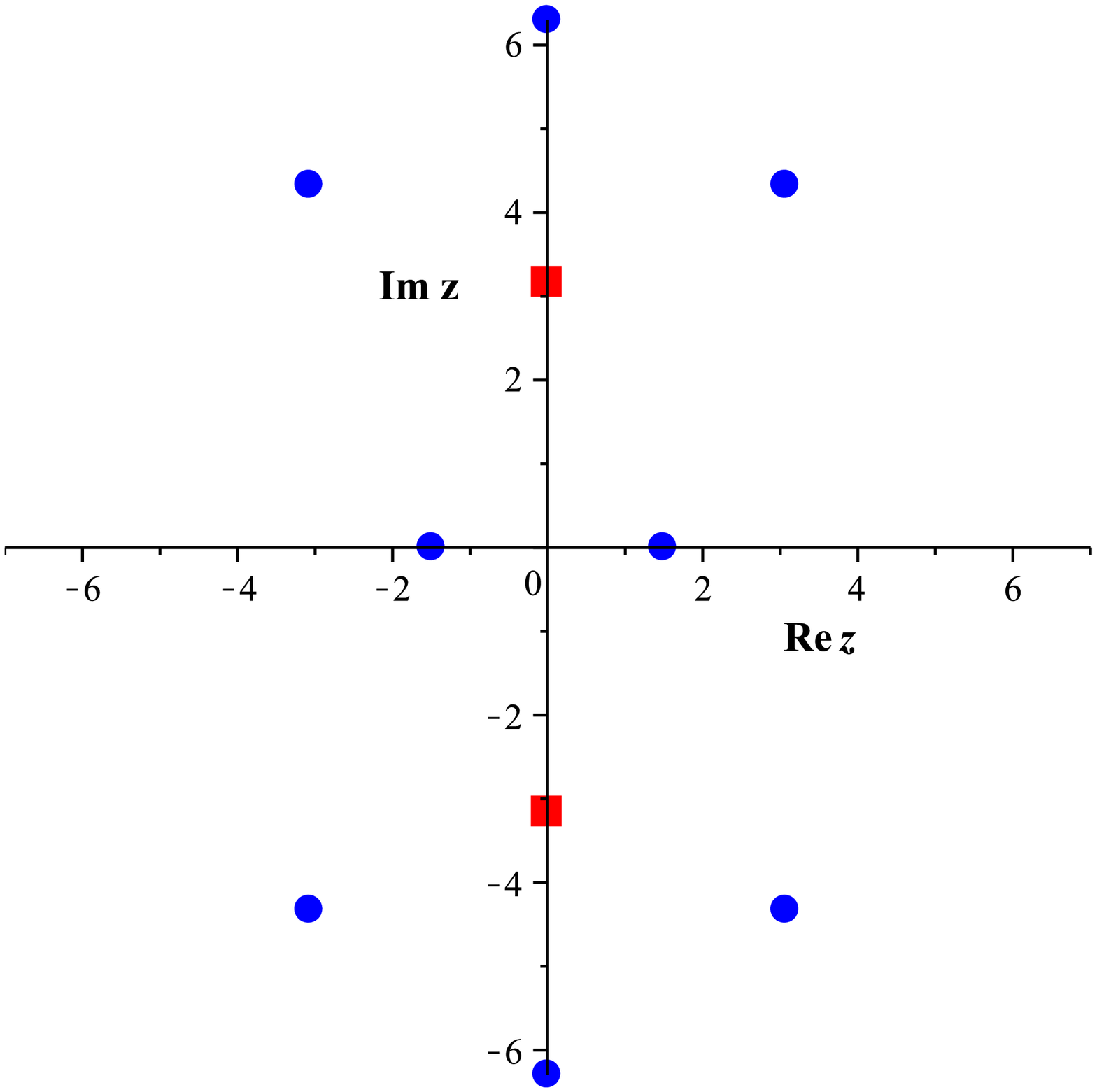,width=60mm}\label{}}
 \subfigure[]{\epsfig{file=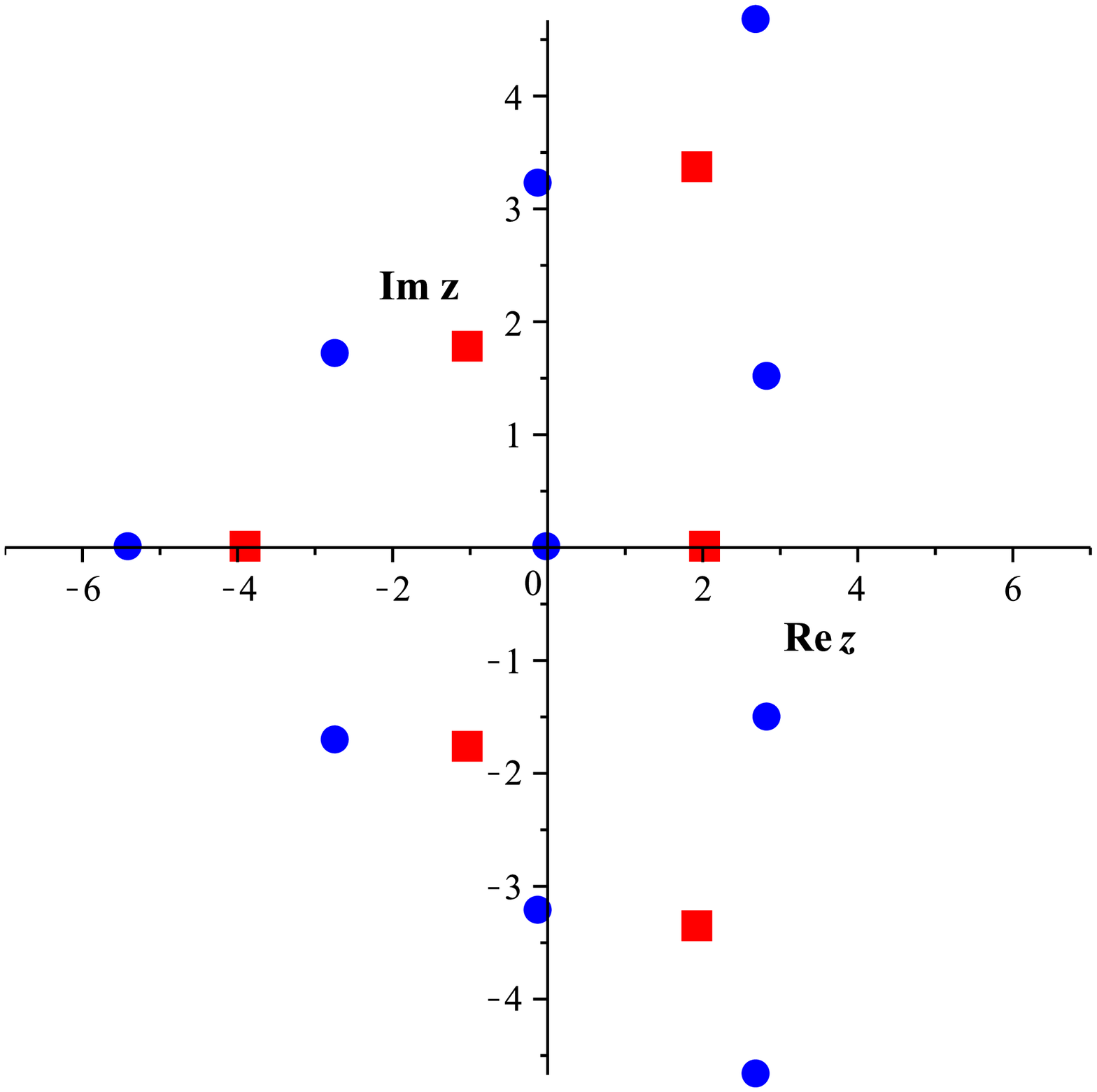,width=60mm}\label{}}}
   \caption{Plots of vortex positions in stationary equilibrium described by polynomials \eqref{Pols_mu_2_Example} with $s_2=1$, $t_2=10$, $s_3=1$ (figure a) and described by polynomials \eqref{Pols_mu_2_Example_AM} with $c_2=10$, $c_3=1$, $c_4=1$ (figure b). Circles mark vortices with circulation $\Gamma$, squares mark vortices with circulation $-2\Gamma$ (figure a) and $-\Gamma$ (figure b)}
 \label{F:Plots_Pos_of_Vort}
\end{figure}

For example, the following polynomials satisfy equation \eqref{Correlation_for_Polynomials_mu_2}:
 \begin{equation}
\begin{gathered}
\label{Pols_mu_2_Example}S_3(z)={z}^{8}+{\frac {28}{5}}\,t_{{2}}{z}^{6}+14\,{t_{{2}}}^{2}{z}^{4}+28\,{
t_{{2}}}^{3}{z}^{2}+s_{{3}}z-7\,{t_{{2}}}^{4},\quad T_2(z)=z^2+t_2.
\end{gathered}
\end{equation}
In this expression $t_2$, $s_3$ are arbitrary constants. Substituting expressions \eqref{Pols_mu_2_Example} into relations given in \eqref{New_Pols_mu_2_EX}, we see that the polynomials
 \begin{equation}
\begin{gathered}
\label{Pols_mu_2_Example_Tilde}\tilde{P}_3^{(3)}(z)={z}^{10}+{\frac {33}{5}}\,t_{{2}}{z}^{8}+{\frac {98}{5}}t_2^2{z}^{6}+42t_2^3{z}^{4}+s_{{3}}{z}^{3}+21t_2^4{z}^
{2}+s_{{3}}t_{{2}}z-7t_2^5,\\
 \tilde{Q}_3^{(3)}(z)={z}^{6}+3t_{{2}}{z}^{4}+3t_2^2{z}^{2}+t_2^3. \hfill
\end{gathered}
\end{equation}
solve equation \eqref{Equation_PQ_1}. Note that these polynomials are not included into the sequence of the Adler --- Moser polynomials. Indeed the Adler -- Moser polynomials of the corresponding degrees are the following
\begin{equation}
\begin{gathered}
\label{Pols_mu_2_Example_AM}V_4(z)={z}^{10}+15\,c_{{2}}{z}^{7}+7\,c_{{3}}{z}^{5}+c_{{4}}{z}^{3}-35\,c_{{2
}}c_{{3}}{z}^{2}+175\,{c_{{2}}}^{3}z+c_{{4}}c_{{2}}-\frac73c_3^2,\\
 V_3(z)={z}^{6}+5\,c_{{2}}{z}^{3}+c_{{3}}z-5\,{c_{{2}}}^{2}, \hfill
\end{gathered}
\end{equation}
where $c_2$, $c_3$, $c_4$ are arbitrary constants. The roots of the polynomials in \eqref{Pols_mu_2_Example} give positions of vortices with circulations $\Gamma$ (roots of $S_3(z)$) and $-2\Gamma$ (roots of $T_2(z)$) in stationary equilibrium whenever the corresponding polynomials do not have multiple and common roots. Similarly, if the polynomials $V_4(z)$, $V_3(z)$ in \eqref{Pols_mu_2_Example_AM} do not possess multiple and common roots, then their roots provide positions of vortices with circulations $\Gamma$ (roots of $V_4(z)$) and~$-\Gamma$ (roots of $V_3(z)$) in stationary equilibrium. Several examples are plotted in figure \ref{F:Plots_Pos_of_Vort}.

Sometimes it is supposed that polynomials from the sequences \eqref{Vortices_Polynomials_Solution_mu} provide unique polynomial solutions of equation \eqref{Correlation_for_Polynomials_mu_2} accurate to the freedom discussed above. Let us consider an example and show that it is not the case. The polynomials
\begin{equation}
\begin{gathered}
\label{Ex_mu_2}S(z)=z^5(z+7\delta)(z^2+9\delta z+21\delta^2),\quad T(z)=z^5(z+7\delta)^2
\end{gathered}
\end{equation}
satisfy equation \eqref{Correlation_for_Polynomials_mu_2}. Studying the structure of these polynomials with the help of the results of section \ref{Stationary_case}, we obtain that they describe equilibria of four point vortices ($\delta\neq 0$) in the following arrangement
\begin{equation}
\begin{gathered}
\label{Ex_mu_2_Circ}\Gamma_1=-5,\quad a_{1}^{(1)}=0,\hfill \\
\Gamma_2=-3,\quad a_{1}^{(2)}=-7\delta,\hfill \\
\Gamma_3=1,\quad a_{1}^{(3)}=\left(-\frac92+\frac{\sqrt{3}}{2}i\right)\delta,\quad a_{2}^{(3)}=\left(-\frac92-\frac{\sqrt{3}}{2}i\right)\delta. \hfill
\end{gathered}
\end{equation}
The polynomials of the corresponding degrees from the sequences \eqref{Vortices_Polynomials_Solution_mu} are
\begin{equation}
\begin{gathered}
\label{Ex_mu_2_Sequences}S_3(z)={z}^{8}+{\frac {28}{5}}\,t_{{2}}{z}^{6}+14\,t_2^{2}{z}^{4}+28\,
t_2^{3}{z}^{2}+s_{{3}}z-7\,t_2^{4}, \hfill \\
 T_4(z)={z}^{7}+7\,t_{{2}}{z}^{5}+35\,{t_{{2}}}^{2}{z}^{3}+t_{{4}}{z}^{2}-35\,
{t_{{2}}}^{3}z-\frac52\,s_{{3}}+t_{{4}}t_{{2}}, \hfill
\end{gathered}
\end{equation}
where $t_2$, $s_3$, $t_4$ are arbitrary constants. We see that the polynomials given by \eqref{Ex_mu_2} can not be included into these sequences. Thus we have constructed an alternative polynomial solution of the equation \eqref{Correlation_for_Polynomials_mu_2}. Note that while comparing the polynomials one should take into account the invariant transformation $z\mapsto \alpha z+\beta$.

Now let us study the generalized Tkachenko equation \eqref{Equation_PQ_1_Translating}. If $\tilde{P}(z)$, $\tilde{Q}(z)$ in \eqref{Equation_PQ_1_Translating} are polynomials, then they have equal degrees. This fact  can be obtained balancing the highest--order terms in equation \eqref{Equation_PQ_1_Translating}.  It is well known that equation \eqref{Equation_PQ_1_Translating} possesses polynomial solutions given by the Adler - Moser and the modified Adler - Moser polynomials (see \cite{Clarkson01}). Let us note that the parameter $\lambda$ can be removed from the equation \eqref{Equation_PQ_1_Translating} if we introduce the new variable $\xi=\lambda z$.

Frequently it is stated that polynomial solutions of equation \eqref{Equation_PQ_1_Translating} have triangular degrees. Let us consider an example and show that this statement is not valid. Suppose we study translating equilibria of two group of point vortices: one with circulation $\Gamma$ and generating polynomial $P(z)$ and another with circulation $-\Gamma$ and generating polynomial $Q(z)$. Then the polynomials $P(z)$, $Q(z)$ satisfy the equation
\begin{equation}
\label{Equation_PQ_Translating_Example}P_{zz}Q-2P_zQ_z+PQ_{zz}-2\lambda_0\left(P_zQ-PQ_z\right)=0,\quad \lambda_0=\frac{\lambda}{\Gamma}.
\end{equation}
In what follows we denote solutions of equation \eqref{Equation_PQ_Translating_Example} as $P(z;\lambda_0)$, $Q(z;\lambda_0)$ According to the formula \eqref{New_Pols_Arbitrary_Circ} we see that the functions
\begin{equation}\begin{gathered}
\label{Translating_Example_Substitution}\tilde{P}(z;\lambda)=P^{\frac{\Gamma(\Gamma+1)}{2}}(z;\lambda_0)Q^{\frac{\Gamma(\Gamma-1)}{2}}(z;\lambda_0),\quad  \lambda_0=\frac{\lambda}{\Gamma} \hfill \\
 \tilde{Q}(z;\lambda)=P^{\frac{\Gamma(\Gamma-1)}{2}}(z;\lambda_0)Q^{\frac{\Gamma(\Gamma+1)}{2}}(z;\lambda_0), \quad \lambda_0=\frac{\lambda}{\Gamma}
\end{gathered}
\end{equation}
solve the generalized Tkachenko equation \eqref{Equation_PQ_1_Translating}.
Thus for any solution $P(z;\lambda_0)$, $Q(z;\lambda_0)$ of equation \eqref{Equation_PQ_Translating_Example} transformation \eqref{Translating_Example_Substitution} gives a solution of the generalized Tkachenko equation \eqref{Equation_PQ_1_Translating}. Let us take $\Gamma=3$. The polynomials
\begin{equation}
\label{Translating_PQ_Example}P(z)=z+\frac{1}{\lambda},\quad Q(z)=z
\end{equation}
provide a solution of the generalized Tkachenko equation \eqref{Equation_PQ_1_Translating}. Using formula \eqref{Translating_Example_Substitution}, we see that the polynomials
\begin{equation}
\label{Translating_PQ_Example_Tilde_SC}\tilde{P}(z)=z^3\left(z+
\frac{3}{\lambda}\right)^6,\quad \tilde{Q}(z)=z^6\left(z+\frac{3}{\lambda}\right)^3.
\end{equation}
solve the same equation. We note that the degree of the polynomials $\tilde{P}(z)$, $\tilde{Q}(z)$ in \eqref{Translating_PQ_Example_Tilde_SC} is not a triangular number.



\section{Conclusion}

In this article we have studied the problem of finding stationary and translating equilibrium positions for multivortex systems with circulations $\Gamma_1$, $\ldots$, $\Gamma_N$. We have found ordinary differential equations satisfied by generating polynomials of vortex arrangements. We have shown that the stationary case can be described with the help of the Tkachenko equation. Strikingly it turns out that the Adler -- Moser polynomials are not unique polynomial solutions of this equation.

We have obtained that the equation for generating polynomials in the translating case is reducible to a generalization of the   Tkachenko equation. Usually it is supposed that polynomial solutions of this equation have degrees that are triangular numbers. We have found alternative polynomial solutions of the latter equation, i.e. polynomial solutions with non--triangular degrees.

\section{Acknowledgements}

This research was partially supported by Federal Target Programm
"Research and Scientific--Pedagogical Personnel of Innovation
in Russian Federation on 2009-–2013" (Contract P1228).

\end{document}